\documentclass[12pt,preprint]{aastex}
\received{2004 June 21}

\newcommand{\lsim}{{\, \lower2truept\hbox{
${< \atop\hbox{\raise4truept\hbox{$\sim$}}}$}\,}}
\newcommand{\gsim}{{\, \lower2truept\hbox{
${> \atop\hbox{\raise4truept\hbox{$\sim$}}}$}\,}}

\begin{document}
%\tighten
\def\etal{{\it et al.\/}}
\def\cf{{\it cf.\/}}
\def\ie{{\it i.e.\/}}
\def\eg{{\it e.g.\/}}
\def\n0{\dot{n}_\circ}
\def\d5{\dot{n}_{\circ,0.5}}

%last revision by D.G. 27/7/2006

\title{On the Rates of Gamma Ray Bursts and Type Ib/c Supernovae}

\author{{\bf Dafne Guetta\altaffilmark{1},
Massimo della Valle\altaffilmark{2}}}

\altaffiltext{1}{INAF-Osservatorio Astronomico di Roma, Monteporzio
Catone (Roma), Italy.} \altaffiltext{2}{INAF-Osservatorio
Astrofisico di Arcetri, Firenze, Italy.}

\begin{abstract}

We measure the local rates of ``low-luminosity'' 
(LL-GRBs, i.e. $L<10^{48\div 49}$erg/sec) and
``high-luminosity'' Gamma-ray Bursts (HL-GRBs). The values are in the
range $\n0=100\div 1800$ Gpc$^{-3}$ yr$^{-1}$ and $\n0=100\div 550$
Gpc$^{-3}$ yr$^{-1}$, respectively, and the ratios to SNe-Ibc 
$\sim 1\%-9\%$ and $0.4\% -3\%$. 
These data may suggest the existence of two physically distinct classes of 
GRBs in which LL-GRBs are (intrinsically) more frequent 
events than  HL-GRBs.
However, with the present data we cannot exclude the possibility 
of a single population of GRBs  which give rise to both an isotropic
low-luminous emission (LL-GRBs: detectable only in nearby GRBs)
and to a highly collimated high-luminous emission (HL-GRBs: detectable
preferentially at high-z).
We  compute also the rate of SNe-Ibc
characterized by broad-lined spectra (Hypernovae) and found it to be about
$1.5\times 10^{-4}$HNe yr$^{-1}$ $10^{10}$ L$_{B\odot}$ 
(i.e less than $10\%$ of SNe-Ibc occurring in Spirals). 
This result implies that the ratio HL-GRBs/HNe is smaller than 1, 
possibly in the range 0.04--0.3. 
We have used the ratio between Hypernovae and LL-GRBs to constrain 
their beaming factor to $f_b^{-1}\sim 10$ or less.
\end{abstract}


\section{Introduction}

Multiwavelenght follow-up studies of GRBs in the last decade have
established that a significant fraction of long GRBs arise from the
simultaneous collapse of a massive star (e.g. Woosley \& Bloom 2006,
Della Valle 2006). Best examples of this are the SN/GRB associations
so far discovered in the local universe, 2006aj with GRB 060218 (Pian
\etal~2006, Campana \etal~2006, Modiaz et al. 2006) and SN 1998bw with
980425 (Galama \etal~1998). Further evidence ($0.1<z<0.2$) comes from
SN 2003dh with GRB 030329 (Stanek \etal~2003; Hjorth \etal~2003) and
GRB 031203, which have been found to be associated with bright
broad-lined type Ic SNe (Malesani \etal~2004). This type of GRB/SN
association applies to a significant fraction of long GRBs, but not
all of them (Della Valle et al. 2006a, Fynbo et al. 2006, Gal-Yam et
al. 2006, Gehrels et al. 2006). At larger redshifts the association
GRB-SNe relies on about a dozen of rebrightenings (``bumps'') observed
during the late decay stages of the GRB afterglow light curve (see Zeh
\etal~2004 and references therein). They have been interpreted as due
to the emergence of the optical contribution of an underlying SN
(Bloom \etal~1999). In two cases, GRB 021211/SN 2002lt (Della Valle
\etal~2003) and GRB 050525A/SN 2005nc (Della Valle \etal~2006b),
spectroscopic observations obtained during these ``bumps'' are
suggestive of the presence of SN components. In spite of these
remarkable achievements, there are still many uncertainties hampering
our knowledge the rate of these events. Particularly, the recent
discovery of GRB 060218 at z=0.03 has raised the question whether or
not a population of ``local'' and ``Low-Luminosity'' GRBs (LL-GRBs,
i.e. $L<10^{48\div 49}$ erg/sec) with different properties from the
energetically ``High-Luminosity'' GRBs (HL-GRBs) does exist (e.g. Cobb
\etal~2006, Soderberg et al. 2006a, Amati et al. 2006).

The aim of this Letter is twofold: first we derive the 
rate for the local ($z\lsim 0.1$) and LL-GRBs, 
with two independent approaches. Then, we investigate the existence 
of two classes of GRBs by measuring the ratios of the rates of LL-GRBs 
and HL-GRBs to SNe-Ibc progenitors. 

\section{Estimate of the Rate of sub-luminous GRBs}

We consider three LL-GRBs, GRB 980425 (z=0.008), GRB
060218 (z=0.03) and GRB 031203 (z=0.105) detected by {\it BeppoSax}
Wide Field Camera (WFC), {\it Swift} Burst Alert Telescope (BAT) and
INTEGRAL respectively and infer from them an empirical rate for
subluminous GRBs. Following Soderberg et al. 2006a we 
estimate the peak flux in the range 1-1000 keV and compare it with
the threshold peak flux calculated by Band (2003, 2006). In this way
we derive the maximum distance ($D_{\rm max}$) the event could be
detected and therefore the maximum volume ($V_{\rm max}$). Considering
the sky coverage ($S_{\rm cov}$) and the number of years of operation
($T$) of each individual the detector the rate of LL-GRBs 
similar to the GRBs given above is:
\begin{equation}
R_{GRB}=\frac{1}{V_{\rm max}}\frac{1}{S_{\rm cov}}\frac{1}{T}
\end{equation}

{\bf GRB 980425} This burst was detected by {\it BeppoSAX} WFC (Pian
\etal~1999) and by BATSE (Kippen \etal 1998). The peak flux in the
50-300 keV band was $F_{50-300}=4.48$ ph cm$^{-2}$ s$^{-1}$.
The redshift of this burst is z=0.0085, implying an isotropic peak luminosity
$L_{980425}\sim 5\times
10^{46}$ erg s$^{-1}$ several orders of magnitude smaller than typical
long HL-GRBs that have $L\sim 10^{51}$ erg$s^{-1}$ on average. By using the
spectrum given by Jimenez, Band \& Piran (2001) we extrapolate the peak
flux $F_{1-1000}=7.6$ ph cm$^{-2}$ s$^{-1}$.  Given the threshold
$F_T=0.8$ ph cm$^{-2}$ s$^{-1}$ we find $D_{\rm max}=120$ Mpc.  The BeppoSax
Sky coverage was about 0.08 and the operation time $\sim$ 4 yrs. Therefore
the rate of GRB 980425-like events is $R\sim 430$ Gpc$^{-3}$
yr$^{-1}$.

{\bf GRB 060218}. This burst was detected by {\it Swift}/BAT. The peak
flux was $F_{15-150}=2\times 10^{-8}$ erg cm$^{-2}$ s$^{-1}$ and the
spectral index  1.5 (Campana \etal~2006). 
The redshift is, z=0.03, implying an isotropic peak luminosity
 of the order of $L_{980425}$.
The extrapolated peak flux is
$F_{1-1000}=1.37$ ph cm$^{-2}$ s$^{-1}$. Given the threshold $F_T=1$
ph cm$^{-2}$ s$^{-1}$ we find $D_{max}=160$ Mpc. The {\it Swift} 
Sky coverage was 0.17 and the operation time is one year. 
Therefore we estimate that the rate of GRB 060218-like events
 is $R\sim 350$ Gpc$^{-3}$ yr$^{-1}$. 

{\bf GRB 031203} This burst was detected by INTEGRAL (Malesani
\etal~2004). The peak flux in the 20-200 keV band was
$F_{20-200}=1.2$ ph cm$^{-2}$ s$^{-1}$ and the spectral index is 0.8
($F_{\nu}=\nu^{-0.8}$). 
The redshift of this burst is, z=0.105, implying an 
isotropic peak luminosity $L_{031203}\sim 3\times
10^{48}$ erg s$^{-1}$ that is three order of magnitude smaller than
canonical long bursts but two order higher than the previous ones. 
The extrapolated flux is $F_{1-1000}=3.35$ ph
cm$^{-2}$ s$^{-1}$. Given the threshold $F_T=0.7$ ph cm$^{-2}$ s$^{-1}$
(Mereghetti \& G\"otz 2005) we find $D_{max}=950$ Mpc. The INTEGRAL 
Sky coverage is 0.5\% and the operation time was 3 yrs. Therefore the rate of
GRB 031203-like events is $R\sim 25$ Gpc$^{-3}$ yr$^{-1}$.
The redshift and the luminosity of this burst are
intermediate between LL-GRBs and HL-GRBs and this suggest a
continuity between the two classes.

The occurrence of two events such as GRB 980425 and GRB 060218, within
$\sim 150$ Mpc distance implies a rate of
$380^{+620}_{-225}$ LL-GRB Gpc$^{-3}$ yr$^{-1}$ in agreement with
Soderberg et al. 2006a result.  The attached errors represent $1\sigma$
Poissonian standard deviation (Gherels 1986). Another potential source
of uncertainty (which is not included) is the correction for
beaming. However, we point out that LL-GRBs may be much less
collimated than typical HL-GRBs (see next section). 
One way to estimate the jet opening angle, $\theta$, is to look 
for ``achromatic-breaks'' in the afterglow light curve.
 The steepening of the afterglow light curve, break, occurs
when the bulk Lorentz factor of the relativistic outflow becomes lower
than $1/\theta$.
From the data on the afterglow of GRB 980425 and GRB 060218 it seems 
that the presence of breaks in the  light curves is very late, 
if existent at all, implying very large $\theta$.

For example radio observations of GRB 060218
(Soderberg \etal~2006a) have  set $\theta \gsim 70^{\circ}$
which corresponds to a beaming factor ($f_b=1/(1-cos\theta)$),
$f_b^{-1}\lsim1.5$. 
By interpreting the break in
the X-ray light curve of GRB 031203 as due to a jet, we obtain $\theta
\sim 30^{\circ}$, which corresponds to $f_b^{-1}\sim 8$.
The beaming corrected rate would thus be $R\sim 25\times 8 \lsim 200$ event
Gpc$^{-3}$ yr$^{-1}$, consistent with the value reported above.

\section{Comparison with the cosmological high luminous
and low-z low-luminous GRB rates}

The above estimate of the LL-GRBs rate may be affected
by the small number of available events, therefore an independent
check is in order. To do this we compare the empirical rate with
the rate inferred from the  large GRB samples obtained by
both BATSE and Swift. As suggested by Schmidt (2001), the GRB local
rate can be estimated by using the BATSE peak flux distribution.
This is given by the convolution of two unknown quantities, the
luminosity function (LF) and the GRB formation rate.  Following
Guetta, Piran \& Waxman (2005) and Guetta et al. (2004), we assume
that GRBs trace the star formation history and adopt the
Rowan-Robinson star formation rate (SFR), $ R_{GRB}(z) = \n0 \, {\rm
min}(10^{0.75},10^{0.75\,z})$, (Rowan-Robinson 1999).  
Other rate evolutions have been analized in detail in Firmiani et al.
2004. Then we find
the best fit to the LF by comparing the BATSE observed peak flux
distribution with the predicted one, assuming a form for the LF
(i.e. a power law with a minimum and maximum luminosities).  We show
in Fig. 1 that  a single power-law LF, $\Phi(L)=L^{-1.6}$, with a
minimum ($L_{\rm min}=5\times 10^{49}$ erg s$^{-1}$) and maximum
luminosity (L$_{\rm max}=5\times 10^{52}$ erg s$^{-1}$) fits the BATSE peak
flux distribution very well.  To obtain the observed local
rate of GRBs per unit volume, $\n0$, we need to estimate the effective
full-sky coverage of our GRB sample.  The BATSE catalog represents
3.185 years of BATSE full sky coverage implying a rate of 692 GRB per
year. Using our LF we find $\n0\sim 1.1
\,$Gpc$^{-3}$yr$^{-1}$, that is much
smaller than what would be required in order to explain the local rate of 
LL-GRBs\footnote{Note that for a different SFR like
for example the SF2-SFR of Porciani and Madau (2001) we still find a
good fit and a local rate smaller by a factor $\sim 2$.}.

However the local rate discussed in the previous paragraph was derived 
under the hypothesis that classical bursts exceed by far the
luminosity of GRB 980425, of GRB 060218 and of GRB 031203 (i.e. in the
estimate of the rate Guetta, Piran \& Waxman (2005)
 considered a minimum luminosity for the
luminosity function (LF) $L_{\rm min}\gtrsim 5\times 10^{49}$ erg/sec):
as such, this rate cannot be compared with the
empirical one derived for LL-GRBs. A self consistent estimate can
only be obtained with a LF that extends down to luminosities as low as that
of GRB 980425 and GRB 060218.

It is important to realize that most of the bursts below $L\sim
10^{48\div 49}$ erg s$^{-1}$ (LL-GRBs) are undetectable by current detectors,
unless they are extremely close (z$<0.1$).  
On the basis of the observables that we have (i.e. the peak flux 
distribution) we cannot constrain the minimum value of the LF, 
therefore we find reasonable to take the minimum luminosity
observed in GRBs ($L_{980425}$) as our minimum value for the LF.

If we repeat the same procedure given above using the same 
LF and L$_{\rm max}$ but a different $L_{\rm min}=L_{980425}$,
 we find that the BATSE peak flux distribution is fitted 
very well and the rate increases as $\n0\sim 200\,$Gpc$^{-3}$yr$^{-1}$.

The increase in the rate of LL-GRB is due to the fact that
the luminosity function increases rapidly with decreasing luminosity
from $5\times 10^{49}$ ergs/sec down to $L_{\rm min}=L_{980425}$
(Guetta \& Piran 2006).
If we further extend the LF considering  $L_{\rm min}$  in the range between
$0.1\,L_{980425}-L_{980425}$ we obtain $\n0=200-1800$
Gpc$^{-3}$ yr$^{-1}$ consistent with the empirical rate determined in
section 2.

This luminosity function fits well also the Swift peak flux
distribution.  Taking into account that Swift has detected 122 long
GRBs in 1.6 yr and covers 1/6 of the sky, applying the same method
described above, we find a rate of $\n0=110-1200$ for $L_{\rm min}$ in
the range between $0.1\,L_{980425}-L_{980425}$ consistent with the
rate derived from the BATSE data and with the empirical rate.

\section{Rate of SNIbc, Hypernovae and GRBs}

The fraction of SNe-Ibc that produces GRBs can be measured as follows.
A rate of $\sim 2 \times 10^4$ SNe-Ibc Gpc$^{-3}$ yr$^{-1}$ is
derived by combining the local density of B luminosity of $\sim
1.2\times 10^8 L_{B,\odot}$ Mpc$^{-3}$ (e.g. Madau, Della Valle \&
Panagia 1998) with the rate of SNe-Ibc observed in Sbc--Irr
Hubble types (these morphological types are appropriate to represent
the GRB hosts) of 0.16 SNe per century and per 10$^{10}$ $L_{B,\odot}$
(SNu units, Cappellaro, Evans \& Turatto 1999). This SN rate has to be
compared with the rate of HL-GRBs of $\sim 1.1$ GRB Gpc$^{-3}$
yr$^{-1}$ (Guetta, Piran \& Waxman 2005) after rescaling for the jet
beaming factor, $f_b$. There exist different estimates for this
parameter: from $\sim 75$ (Guetta, Piran \& Waxman 2005) to $\sim 500$
(Frail \etal~2001) corresponding to beaming angles $\sim
10^\circ$--$4^\circ$, respectively. Taking these figures at their face
value, we find the ratio HL-GRB/SNe-Ibc to be in the range: $\sim
0.4\%-3\%$ and the ratio LL-GRBs/SNe-Ibc: $\sim 1\%-9\%$ (for
$f_b^{-1}=1$). Radio surveys give  independent and consistent
constraints: Berger \etal~(2003) find that the incidence of SN
1998bw-like events, in the nearby universe, is $\lsim 3\%$, 
Soderberg et al. (2006c) find HL-GRB/SNe-Ibc $<10\%$.

The computation of the GRB/HNe ratio requires a further step. The
measurement of the SN rate is based on the control-time methodology
(Zwicky 1938) that implies the systematic monitoring of galaxies of
known distances and the use of appropriate templates for the light
curves of each SN type (see Cappellaro \etal~1993 for bias and
uncertainties connected with this procedure). Unfortunately all
Hypernovae reported in Tab. I have not been discovered during time
`controlled' surveys, and therefore any attempt to derive an absolute
value of the rate of Hypernovae should be taken with caution.  One
possibility is to compute the frequency of occurrence of all SNe-Ib/c
and Hypernovae in a limited distance sample of objects.  We extracted
193 SNe-Ib/c from an upgraded version of the Asiago catalog
(\texttt{http://web.pd.astro.it/supern}), 19 of these have been
spectroscopically classified as Hypernovae. 

%In Fig. 1 we report the cumulative distributions of the recession
%velocities of the hosts of HNe (i.e. broad-lined SNe-Ib/c) and
%``standard'' SNe-Ibc (dotted line).  A K-S test shows that the
%probability that the two SN samples originate from the same population
%is $\sim 9\%$. This result may be related to an observational bias due
%to the epoch in which the spectroscopic classification, in either
%``standard'' or ``broad-lined'' SN-Ibc, was carried out. Indeed even
%if the pre-maximum spectra of HNe show significantly broader lines
%than ``standard'' Ibc, this difference vanishes after maximum, such
%that it may not be easy to distinguish between the two types of SNe.
%However this behavior almost disappears if we consider all SNe within

In Fig. 2 we show that the cumulative distributions of the recession
velocities of the hosts of HNe (i.e. broad-lined SNe-Ib/c) and
``standard'' SNe-Ibc (dotted line) are statistically indistinguishable
(KS probability=0.73) for $cz< 10,000$ km/s.  Within this volume we
find 158 objects, 12 of which are HNe. After excluding SN 1998bw
and 2006aj, because they were not serendipitously discovered and
assuming that the host galaxies of both `normal' SNe Ib/c and
Hypernovae were monitored with a comparable level of efficiency, we
infer that the fraction of Hypernovae is about $10/156 \simeq 7\%$ of
the total number of SNe Ib/c. Therefore the HN rate turns out to be
$\sim$ 0.015 SNe per century and per 10$^{10}$ $L_{B,\odot}$. 
Note that this value is about an order of magnitude larger than
reported by Podsiadlowski et al. (2004). These authors find a similar
ratio HNe/SNe-Ibc (about 5\%), however they assume that HNe are on
average 4 times brighter than ``standard'' SNe-Ibc, and therefore they
can be detected in a volume about 10 times larger, therefore they
adopt an ``actual'' ratio HNe/SNe-Ibc $\sim 0.5\%$. However their
assumption on the luminosity at maximum of HNe does not seem supported
by either our Fig. 2 and the results of Soderberg et al. (2006b) who
find the magnitude distribution of GRB/XRF-SNe and local SNe-Ib/c (see
Richardson et al. 2006) to be statistically indistinguishable.

Finally we check whether our result is consistent with the lack of a
single HN detection in the Cappellaro \etal's~(1999) sample. In
particular, from Poisson statistics we find that the
probabilities of obtaining a null result are $\sim$50\% and $\sim$5\% when the
expected HN numbers are 0.7 and 3.0. After using the control time of
Cappellaro \etal~(1999) we derive $\sim 0.012$ and $0.05$ HNe per
century per 10$^{10}$ $L_{B,\odot}$. Then, one should expect that the
HN rate is likely similar to the 50\% probability value (0.012 SNu)
and hardly higher than the 5\% probability value (0.05 SNu). The rate
of 0.015 SNu derived in this paper is fully consistent with the limits 
set by the poissonian statistics.  
Together these data imply a ratio LL-GRBs/HNe in the range
$\sim 1\%-10\%$ and beaming factors $f_b^{-1}\lsim 10$ (or $\theta \gsim
25^\circ$). In view of these findings, we shall assume, in the following,
that LL-GRBs emit (almost) isotropically.

It is interesting to compare the beaming
corrected rate of  HL-GRBs with the  rate of LL-GRBs. For every
observed HL-GRB there are $f_b^{-1}$ bursts that are not observed,
thus the true rate is $R\sim 100-550$ Gpc$^{-3}$
yr$^{-1}$. For isotropic LL-GRBs our analysis yields comparable rates:
$R\lsim 150-600$ Gpc$^{-3}$ yr$^{-1}$ (see section 2), $R\lsim
100-1200$ Gpc$^{-3}$ yr$^{-1}$ (from Swift data, section 3) and
$R\lsim 200-1800$ Gpc$^{-3}$ yr$^{-1}$ (from BATSE data, section
3). These results indicate that if there are two populations of GRBs
(HL-GRBs and LL-GRBs) 
the respective frequency of occurrence are 
comparable within a  factor of $\sim 3$. However the data do not exclude the
case that only one population of GRBs is responsible for 
giving rise to  both the highly collimated component (preferentially
observable at high z) and to the almost isotropic components (detectable
only in nearby GRBs) (e.g. Woosley \& Heger 2006).

\begin{table}
\begin{tabular}{crr}
\hline
  \tablehead{1}{r}{b}{SN}
& \tablehead{1}{r}{b}{cz~km/s}
& \tablehead{1}{r}{b}{References}\\
\hline
     1997dq & 958 &  Mazzali \etal~2004\\
     1997ef & 3539&  Filippenko 1997\\
     1998bw & 2550&  Galama \etal~1998 \\
     1999as & 36000& Hatano \etal~2001\\
     2002ap & 632 &  Mazzali \etal~2002, Foley \etal~2003\\
     2002bl & 4757&  Filippenko \etal~2002\\
     2003bg & 1320&  Filippenko \& Chornack 2003a\\
     2003dh & 46000& Stanek \etal~2003, Hjorth \etal~2003\\
     2003jd & 5635&  Filippenko \etal~2003b; Matheson \etal~2003\\
     2003lw & 30000& Malesani \etal~2004\\
     2004af & 16800& Riello \etal~2004\\
     2004bu & 5549&  Foley \etal~2004 \\
     2005da & 4495&  Modjaz \etal~2005\\
     2005fk & 63500& Frieman \etal~2005a\\
     2005kr & 36484& Frieman \etal~2005b\\
     2005ks & 28500& Frieman \etal~2005b\\
     2005kz & 8117&  Filippenko, Foley \& Matheson 2005\\
     2005nb & 7127&  Roman \& Rostopchin 2006\\
     2006aj & 9000&  Masetti \etal~2006\\
    \hline
  \end{tabular}
  \caption{Hypernovae}
\end{table}

%\begin{figure*}[h]
%\centering
%\includegraphics[width=13cm,angle=-90]{f2.ps}
%\caption{\label{all} Cumulative distribution of the radial
%velocities of the hosts of ``standard'' SN-Ibc (solid-line) and HNe
%(dashed-line) within $cz < 9500$ km/s.  A K-S test shows that the
%probability that the two SN samples originate form the same
%population is $\sim 73\%$. }
%\end{figure*}

\begin{figure*}[h]
\centering
\includegraphics{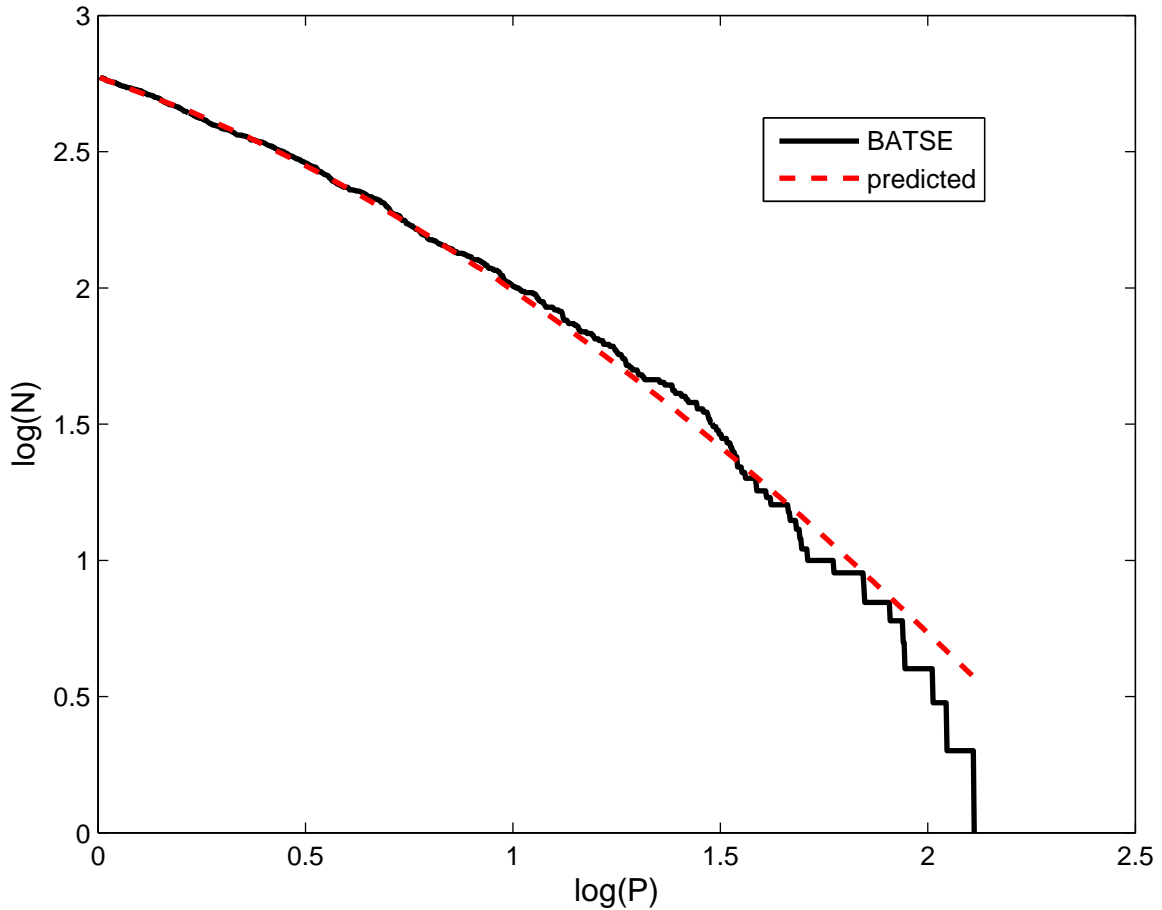}
\caption{\label{all} 
Predicted logN-logP distribution (dashed-line) for the best fit LF with a
Rowan-Robinson SFR (1999) vs. the observed logN-logP taken from the BATSE 
catalog. }
\end{figure*}

\begin{figure*}[h]
\centering
\includegraphics[width=13cm,angle=-90]{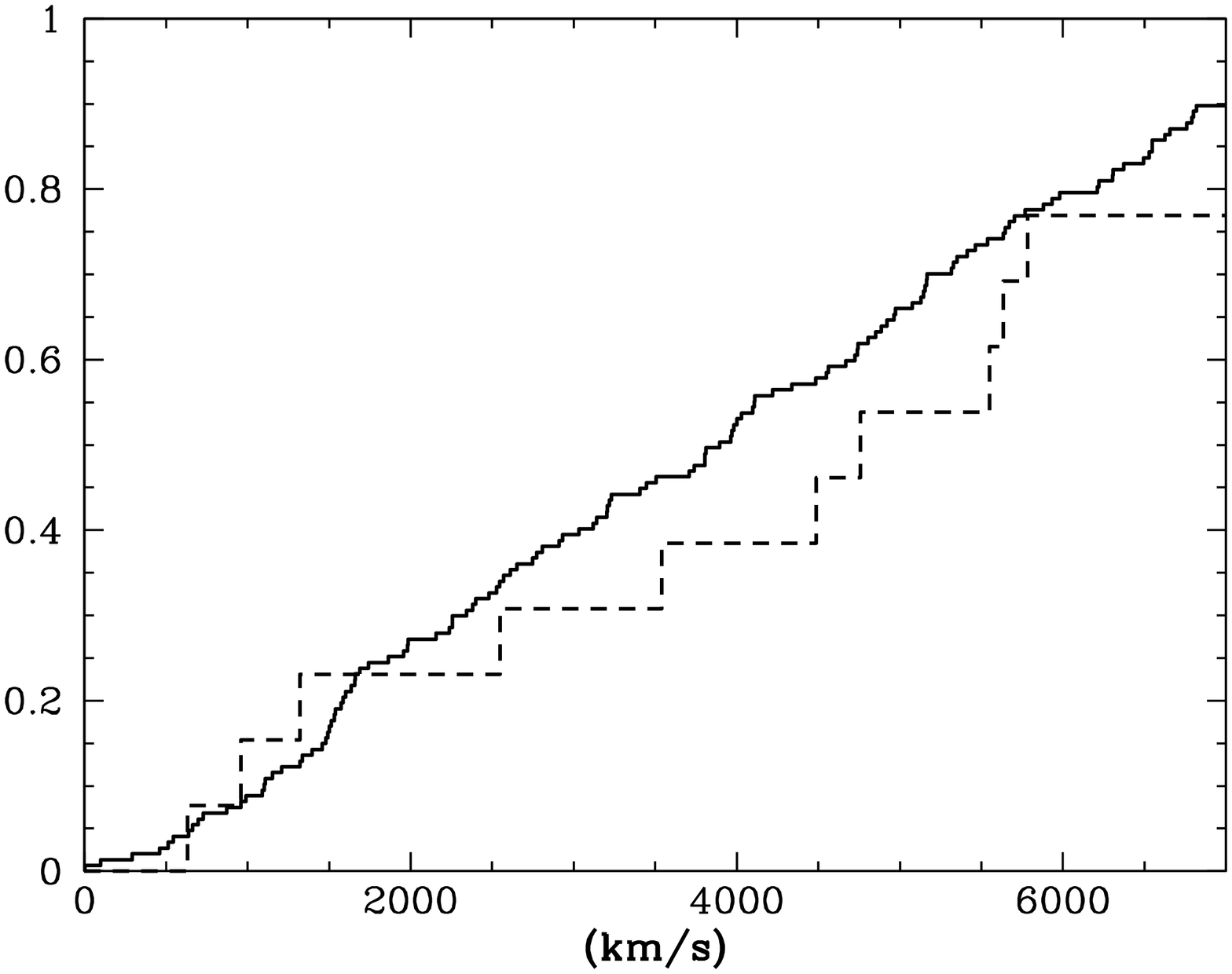}
\caption{\label{all} Cumulative distribution of the radial
velocities of the hosts of ``standard'' SN-Ibc (solid-line) and HNe
(dashed-line), within $cz < 10000$ km/s,
extracted from an upgraded version of the Asiago SN
Catalog (\texttt{http://web.pd.astro.it/supern}). A K-S test shows
that the probability that the two SN samples originate form the same
population is $\sim 73\%$.}
\end{figure*}

\section{Conclusions}

From the available information on GRBs and SNe-Ibc rates a
number of interesting results emerge:

i) we have computed the HN rate, on robust empirical grounds, and
found that this sub-class of SNe-Ibc, characterized by broad-lined
spectra, includes about 7\% of SNe-Ibc. An analysis of the
cumulative distribution of the recession velocities of the respective
hosts, does not suggest that HNe are intrinsically more luminous (on
average) than standard SNe-Ibc. All together these facts imply a
HN rate of $1.5\times 10^{-4}$ HNe $yr^{-1} 10^{10}L_{B\odot}$. 

ii) the ratio HL-GRBs/HNe is smaller than 1, possibly in the range
0.04--0.3. The ratio LL-GRB/HNe is in the range 0.1--1.

iii) if we assume that all HNe are able to produce LL-GRBs, then the
LL-GRB/HNe ratio allows us to constrain the beaming factor to be
smaller than $f_b^{-1}\lsim 10$.

iv) from the analysis of two nearest GRBs we have derived a
LL-GRB rate of $380^{+620}_{-225}$ GRBs Gpc$^{-3}$ yr$^{-1}$. The
attached errors represent $1\sigma$ Poissonian standard
deviation. This result is consistent with the rates of 200--1800 and
100--1200 LL-GRB Gpc$^{-3}$ yr$^{-1}$, derived from the BATSE and
Swift GRBs.

v) the frequencies of occurrence of HL-GRBs (HL-GRB/SNe-Ibc $\lsim
3\%$) and LL-GRBs (LL-GRB/SNe-Ibc $\lsim 9\%$) may suggest the
existence of two physically distinct classes of GRBs (e.g. Cobb et
al. 2006, Soderberg et al. 2006a, Pian et al. 2006, Amati et al. 2006)
in which LL-GRBs are (intrinsically) more frequent events than
HL-GRBs.  However, due to the uncertainties, we cannot exclude that a
single population of GRBs originates both the isotropic and
sub-energetic component, detectable only in nearby GRBs, and the
highly collimated component, observable by sampling huge volumes of
space, and thus mainly detectable in high-z GRBs.

\section{Acknowledgements}

We thank the anonymous referee for useful comments which have improved
the presentation of this paper and Sandro Mereghetti, Daniele Malesani
and Lorenzo Amati for useful discussions. The authors are also
indebted with Luigi Stella and Mario Vietri for their critical reading
of the manuscript. M.D.V. is grateful to the University of Tokyo for
the friendly hospitality and creative atmosphere.  MDV acknowledges
the PRIN-INAF 2005 grant for financial support.

\end{document}